\documentclass[aps,prl,twocolumn,superscriptaddress,floatfix,longbibliography]{revtex4-1}

\usepackage{amsmath,amssymb,verbatim}
\usepackage{physics}
\usepackage{subfig}
\usepackage{wrapfig}
\usepackage{graphicx}
\usepackage{color}
\usepackage{bm}
\usepackage{textcomp} 
\usepackage{enumitem}
\setlist{noitemsep,leftmargin=*}
\usepackage{placeins}
 
\usepackage{hyperref}
\hypersetup{
colorlinks=true,
urlcolor= blue,
citecolor=blue,
linkcolor= blue,
bookmarks=true,
bookmarksopen=false,
}

\begin{document}

\title{Scheme for Majorana Manipulation Using Magnetic Force Microscopy}

\author{Benjamin H. November}
\affiliation{Department of Physics, Harvard University, Cambridge, Massachusetts 02138, USA}

\author{Jay D. Sau}
\email[]{jaydsau@umd.edu}
\affiliation{Condensed Matter Theory Center and Joint Quantum Institute, Department of Physics, University of Maryland, College Park, Maryland 20742, USA}

\author{James R. Williams}
\affiliation{Joint Quantum institute, Center for Nanophysics and Advanced Materials, Department of Physics, University of Maryland, College Park, Maryland 20742, USA}

\author{Jennifer E. Hoffman}
\email[]{jhoffman@physics.harvard.edu}
\affiliation{Department of Physics, Harvard University, Cambridge, Massachusetts 02138, USA}
\affiliation{School of Engineering \& Applied Sciences, Harvard University, Cambridge, Massachusetts 02138, USA}

\date{\today}

\begin{abstract}
We propose a scheme for the use of magnetic force microscopy to manipulate Majorana zero modes emergent in vortex cores of topological superconductors in the Fe(Se,Te) family. We calculate the pinning forces necessary to drag two vortices together and the resulting change in current and charge density of the composite fermion. A possible algorithm for measuring and altering Majorana pair parity is demonstrated.
\end{abstract}

\maketitle
Robust quantum computing is one of the most sought-after breakthroughs in modern science. The primary challenge to completing a useful algorithm is to develop qubits with sufficient immunity to environmental perturbation. There has been a strong push for the investigation of topological quantum computation (TQC) where protection from local fluctuations derives from the non-local nature of the qubit states, greatly increasing their stability \cite{KitaevAnnPhys2003}. Majorana fermions, emergent excitations that satisfy non-abelian exchange statistics, have been suggested as components of a topological qubit.
However, success of this approach requires a method to store and read out the information in the phase of a pair of Majoranas after an exchange.
When two Majoranas are contracted to single point, they can either annihilate or form a canonical fermion. The quantum information stored in the pair is the presence or absence of the charged fermion, and thus measurement of the increased charge or current density constitutes readout of the parity of the Majorana pair. It is thus possible to use the unique braiding statistics and non-locality of Majorana states to create perturbation-resistant qubits.
\par

Majoranas have been heavily studied in semiconductor-superconductor nanowire systems \cite{AlbrechtNat2016,SuominenPRL2017, ZhangNat2018}. Despite hard-won progress in such constructed Majorana environments, evidence for a functional Majorana qubit is still lacking, and there is even residual confusion about whether Majorana modes have been realized at all \cite{LiuPRB2017}. Furthermore, relatively high magnetic fields are generally required in semiconductor nanowire systems to produce the desired topological properties, which can lead to defect decoherence that destroys the Majorana states \cite{PotterPRB2011}. The second challenge is the fine calibration of the chemical potential necessary to achieve the topological phase in a qubit configuration.
These challenges have slowed the practical manifestation of TQC, and thus necessitate a search for topological superconductors (TSCs) of another kind. 
\par

In contrast to the semiconductor-superconductor heterostructures, there are systems that natively exhibit topological superconductivity. These intrinsic TSCs are characterized by the emergence of Majorana bound states at zero energy in vortex cores under an applied magnetic field \cite{FuPRL2008}. The main advantages of intrinsic TSCs are the simplicity and scalability. The planar geometry of intrinsic TSCs is ideal for surface state measurement instruments such as scanning tunneling microscopy (STM) and magnetic force microscopy (MFM). Additionally, the lack of a heterotructure eliminates the need to analyze and perfect of complicated interfaces. As for the scalability of TQC, the number of vortex cores scales simply with the area of the intrinsic TSC and with the applied magnetic field, allowing for easier tuning of the Majorana density. However, the relatively modest magnetic fields required in the intrinsic TSC geometry are less problematic for quasiparticle poisoning than the higher fields in nanowire geometry.
These numerous advantages point to intrinsic TSCs as an ideal platform for the experimental realization of TQC.
\par

Several Fe-based superconductors have been shown to inherently support topological surface states characteristic of a three dimensional topological insulator (TI), in addition to a superconducting gap \cite{XuPRL2016, ZhangNatPhys2019}. The gap has been measured to be $\Delta = 1.8$ meV, comparable to the Fermi energy of $E_F = 4.4$ meV relative to the Dirac point \cite{ZhangSci2018}, making it likely that the subgap states of the vortices, which scale as $\Delta^2/E_F$ \cite{CaroliJPL1964}, are far from the center of the superconducting gap. This has allowed the prediction and observation of topological superconductivity and MZMs in vortices several Fe-based superconductors, by angle-resolved photo emission spectroscopy (ARPES) and STM \cite{LiuPRX2018, KonigArX2019, HaoArX2018}. Specifically, the simple Fe(Se,Te) family has been seen to host MZMs in vortex cores \cite{WangSci2018, MachidaArX2018} and thus has been recognized as a promising candidate for TQC.  
\par

\begin{figure}[h]
\begin{center}
\includegraphics[clip=true,width=\columnwidth]{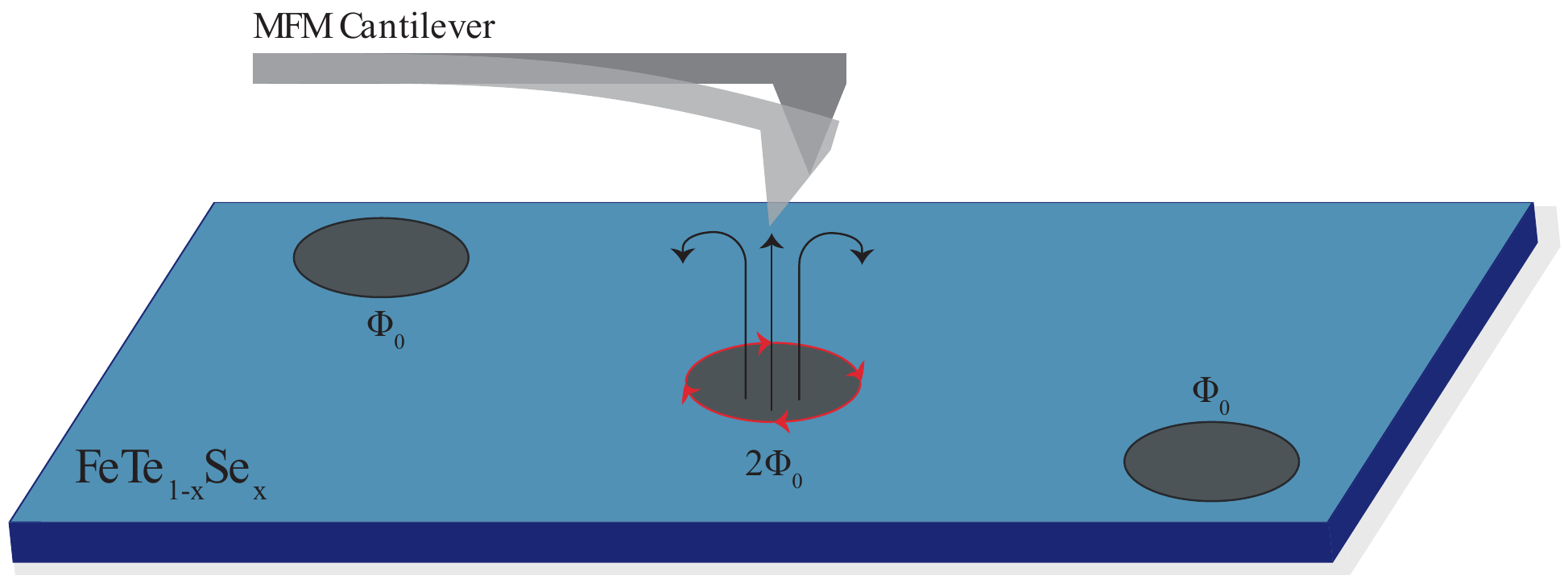}
\end{center}
\caption{MFM cantilever tip measuring the magnetic force generated by the excess supercurrent when a vortex pair with parity 1 is fused into a double vortex.}
\label{fig:MFM-conceptual}
\end{figure}

Here we develop a scheme for braiding Majoranas in FeTe$_{1-x}$Se$_x$ via vortex manipulation by MFM. We propose the use of a cantilever-based tip
that can first detect the MZMs in vortex cores by tunneling, then manipulate the Majoranas by magnetic force. By using a large tip-sample separation during vortex manipulation to avoid quasiparticle poisoning, then approaching the tip rapidly towards a pair of assembled MZMs, we quantify the feasibility to detect the resultant Majorana parity by magnetic force readout, as shown in Fig.~\ref{fig:MFM-conceptual}. By repeated measurements, the Majorana parity lifetime can be measured, and ultimately more complex braiding can be used to conduct logical operations.

\par 
The bulk superconductivity in FeTe$_{1-x}$Se$_x$ has been shown to originate from an electron-like pocket at the M point and a hole-like pocket at the $\Gamma$ point\cite{ZhangSci2018}. Meanwhile, band inversion occurs between an odd-parity $p$ band and an even parity $d$ band along the $\Gamma$-$Z$ direction in plane. The resulting Dirac cone is sharply visible, suggesting negligible scattering of surface states. However, Cooper pairs are free to tunnel between any of the pockets in the Brillouin zone to the surface states, leading to an effective internal proximity effect from the bulk into the topological surface states. Therefore, we consider the FeTe$_{1-x}$Se$_x$ system as realizing the Fu and Kane model for proximity $s$-wave superconductivity on TI surface states \cite{FuPRL2008}.

\par 
We seek to address how to measure the quantum state (fusion channel) of MZMs in a pair of vortices. Each Majorana operator $\gamma_{j=\text{1,2}}$ associated with these MZMs can be considered half a fermionic operator such that the combination $c^{\dagger} = \gamma_1 + i\gamma_2$ is a canonical Fermion creation operator. During a fusion process, the pair of MZMs are brought together to within a coherence length so that the vortices overlap. The quantum state of the vortex pair is then related to the fermionic occupation $c^{\dagger}c$ of the fused vortex pair. For simplicity, in this work we will consider a case where a pair of vortices are brought to completely overlap so that they fuse into a double vortex. Assuming the gap of the resulting double vortex is larger than the thermal smearing, the quantum state of the fermionic operator $c^{\dagger}c$ can be measured from the occupancy of the lowest bound state of the double vortex. 

\par 
The double vortex state can be described by the Bogoliubov-de-Gennes (BdG) Hamiltonian of an $n$-vortex on the surface of a TI:
\begin{equation}
H=[v \boldsymbol{\sigma} \cdot \boldsymbol{p}-\mu]\tau_z+[\Delta(\mathbf{r})\tau_++h.c],
\end{equation}
where $\boldsymbol{p}$ is the momentum operator and $\boldsymbol{\tau}$ and $\boldsymbol{\sigma}$ are the Pauli matrices in Nambu and spin space, respectively \cite{FuPRL2008, JackiwNucPhysB1981}. The gap is taken to be of the conventional form $\Delta(\boldsymbol{r}) = \Delta(r)e^{in\theta} = \Delta_0 \tanh(r)e^{in\theta}$.  In polar coordinates the Hamiltonian becomes
\begin{align}
    H&=[-i v\{(\sigma_x\cos{\theta}+\sigma_y\sin{\theta})\partial_r+\frac{1}{r}(\sigma_y\cos{\theta}-\sigma_x\sin{\theta})\partial_\theta\}]\tau_z\nonumber\\
&-\mu\tau_z+\Delta(r)\{\cos{n\theta}\tau_x+\sin{n\theta}\tau_y\}.
\label{polBdG}
\end{align}
The above Hamiltonian commutes with the total angular momentum operator and can thus be reduced to a radial form by the appropriate choice of unitary transformation. The radial BdG equation then takes the form
\begin{align}
&[-i v\sigma_x\tau_z\partial_r+\tau_z\{-\mu-i (\sigma_y) r^{-1}(-i/2)(\sigma_z+n\tau_z-2m)\}\nonumber \\
&+\Delta(r)\tau_x-E]\Psi(r)=0.
\label{radBdG}
\end{align}
We choose to work in units where $v=\Delta_0=1$, forcing $\xi = 1$ and $\Delta(r)=\tanh(r)$. The BdG can also be made real via the rotation $\sigma_x\rightarrow\sigma_y$ about the $z$ axis. With these choices the BdG equation becomes real and 1-D, given by
\begin{align}
    &[\partial_r+\{-i\mu\sigma_y+ r^{-1}(1+\sigma_z(n\tau_z-2m))/2\} \nonumber \\ 
    &-\tanh(r)\tau_y\sigma_y-i E\sigma_y\tau_z]\Psi(r)=0.
    \label{Psi}
\end{align}

We calculate the solution to the above equation for the $m = \frac{1}{2}$ angular momentum channel of a double vortex ($n=2$). As previously mentioned, the occupation of the state $\Psi(r)$ directly translates to the readout of the vortex pair parity. Thus, measurability of the resulting charge and current densities associated with the fused double vortex wavefunction is crucial to quantum state readout. Since the generated excess charge density is likely screened by the background superconductor, we consider measuring the generated supercurrent around the double vortex using the applied force on an MFM tip. The supercurrent associated with $\Psi(r)$ is given by the expectation value of the tangential component of the current operator defined as
\begin{align}
&j(r)=-v\Psi^{\dagger}(r)\sigma_x\Psi(r).
\end{align}

\begin{figure}[h]
     \includegraphics[clip=true,width=\columnwidth]{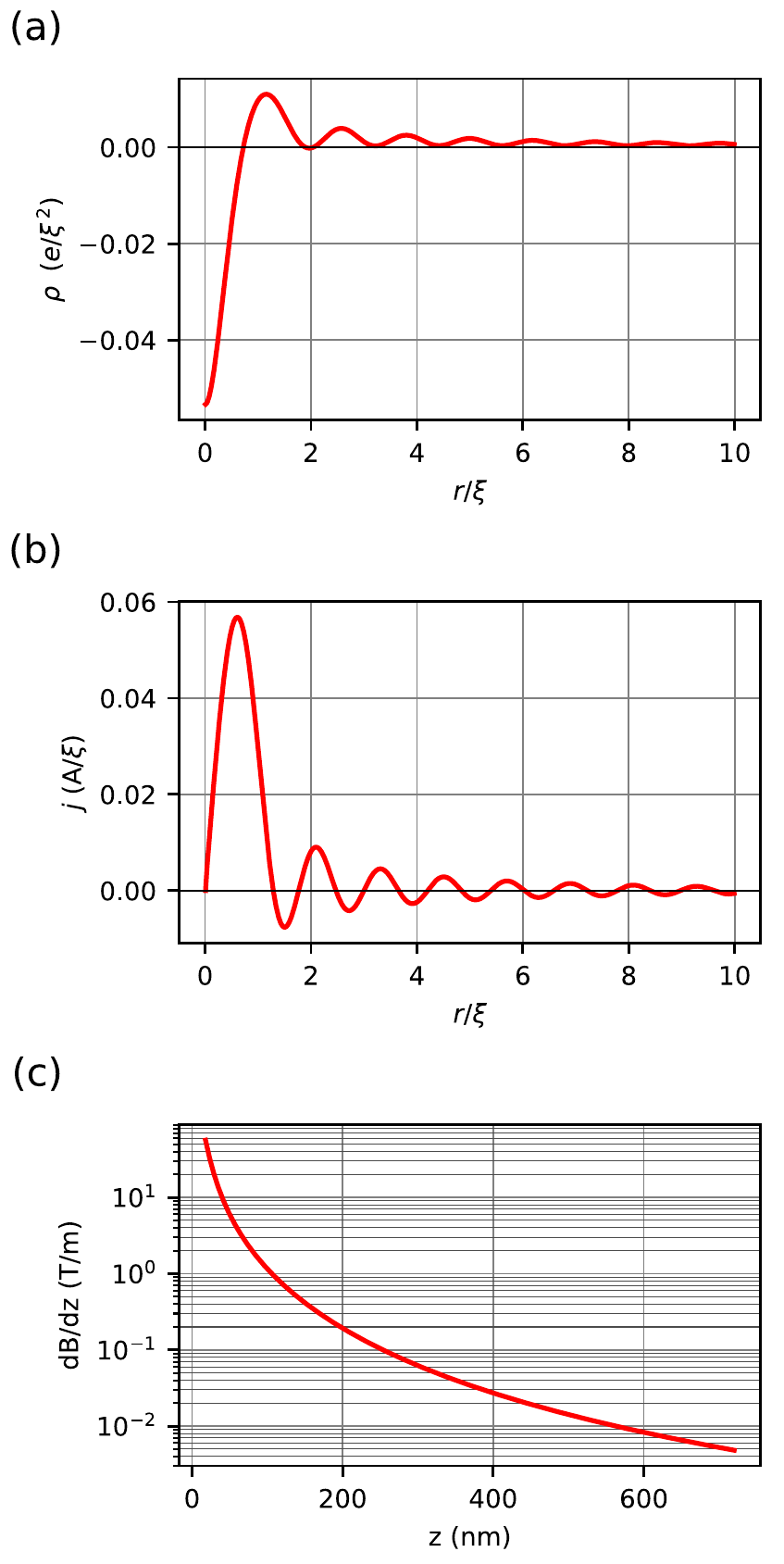}
    \caption{Excess charge (a) and current (b) density around a fused vortex pair with non-zero fermionic occupation. (c) Change in magnetic field gradient as a function of height above the center of the double vortex. Magnetic gradient is larger than 1 T/m up to $\sim100$ nm above the surface of the sample.}
    \label{fig:calc_results}
\end{figure}

Upon approach to the sample above the vortex, the tip will feel a force due to the magnetic field induced above the surface by the excess supercurrent density. Modeling the tip as a magnetic monopole, the change in cantilever frequency can be shown to be
\begin{align}
    \Delta f &= \widetilde{m}\frac{f_0}{2\text{k}}\frac{dB}{dz},
\end{align}
where $k$ is the cantilever force constant, $f_0$ is the resonance frequency, and $\widetilde{m}$ is an effective tip magnetization, typically on order $10^{-8}$ N/T \cite{StraverAPL2008}. The calculated magnetic field gradient, shown in Fig.~\ref{fig:calc_results}, is of order one T/m or larger up to 100 nm away from the surface, well within the working distance of MFM. In order for the force to be measurable, it should produce a change in cantilever frequency on order $\Delta f \sim 10^{-3}$ Hz. Indeed, there are commercially available cantilevers with $f_0=18$ kHz and $k=.06$ N/m for which the excess magnetic field is detectable.

Regardless of efforts to prevent decoherence, there will be a finite lifetime to the topological qubit that is essential to quantify. Ideally, the topological protection of the Majoranas would lead to very long lifetimes. However, the interactions between the Majoranas and stray quasiparticles can cause the parity of qubit pairs to flip, effectively destroying the stored information \cite{CatelaniPRL2011}. In general, parity lifetimes are very difficult to model and are not well understood, although lifetimes on the order of minutes have been realized in superconducting devices \cite{AlbrechtPRL2017}. While quasiparticle poisoning is expected to be a limiting factor in the performance of the Majorana qubit, particularly in a disordered time-reversal broken superconductor, the evidence of conductance quantization and relatively hard gap suggests that the time-scale for such processes might be feasible.

\par
For preliminary experimental studies, quasiparticle poisoning might actually aid in validation of the Majorana measurement process. Although the supercurrent from the occupation of $\Psi(r)$ is likely a small fraction of the total force from the double vortex line, quasiparticle poisoning will lead to a fluctuation of the occupation of the double vortex low-energy bound state that should lead to a characteristic telegraph noise.
This telegraph noise should be measurable as long as the quasiparticle poisoning is within the time-resolution of the force measurement. Observation of such a telegraph noise would be an indication that the fluctuation is indeed Majorana in nature. In addition, such a signature would yield an estimate of the quasiparticle poisoning time-scale that would limit the lifetime of the qubit.

\begin{figure}[h]
    \includegraphics[clip=true,width=\columnwidth]{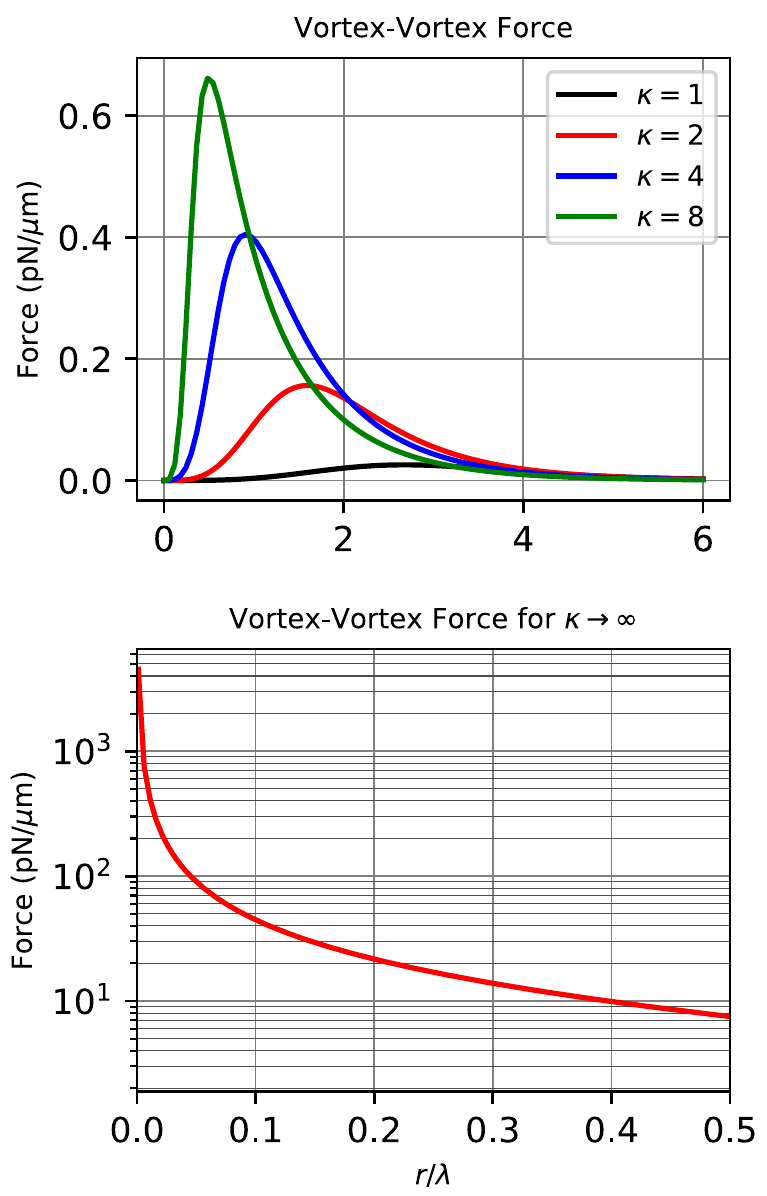}
    \caption{(a) Inter-vortex repulsion force for increasing values of $\kappa$. As the system approaches the type-II limit, the repulsion force becomes a highly peaked, short range interaction. (b) The infinite $\kappa$ behavior of the vortex-vortex repulsion. This extreme type-II limit well approximates the FeTe$_{1-x}$Se$_x$ system.}
    \label{fig:Pinning}
\end{figure}

To braid Majoranas, we will need to move vortices easily around each other. But to read Majorana parity, we will need to bring two vortices together so their cores almost touch, overcoming the inter-vortex repulsion. In Fe(Te,Se), which is at the extreme type-II limit, the vortex-vortex repulsion approaches 300 pN/$\mu$m as the separation approaches 2$\xi$ \cite{ChavesPRB2011}, as shown in Fig.~\ref{fig:Pinning}. Therefore, we must strongly pin one vortex while we bring another vortex towards it.

\par
Previous MFM measurements using Si cantilever and optical detection have manipulated and quantified vortex pinning forces in Nb (detecting vortex jumps smaller than 10 nm) \cite{StraverAPL2008}, cuprates (measuring pinning forces as small as 2 pN/$\mu$m \cite{AuslaenderNatPhys2008}, and Fe-based superconductors (achieving 500 fN resolution of a 4 pN force) \cite{ZhangPRB2015}. For MZM detection purposes, we suggest the use of tuning fork force cantilever, because higher spring constant $k$ and smaller amplitude noise allows simultaneous STM spectroscopy. Switching to tuning fork cantilevers typically increases $k$ from ~ 3 N/m to $\sim3000$ N/m, so it is necessary to ensure that $Q$ and $\omega$ increase by the same factor to avoid reduced sensitivity. In situ feedback can be used to increase $Q$ by a factor of up to 20 \cite{HuefnerAPL2012}. Indeed, low-$T$ MFM with tuning fork has previously shown force resolution to 2 pN, with 15 nm spatial resolution \cite{SeoAPL2005}.

\par
A promising measurement in ion-irradiated Fe(Te,Se) showed that most vortices can be fixed by collective pinning in relatively clean areas, apparently avoiding poisoning of the MZMs by normal quasiparticles present at the pinning site, and leaving sharp zero bias conductance peak (ZBCP) intact \cite{MasseeSciAdv2015}. However, recent STM studies in Fe(Te,Se) have shown that the likelihood of finding a ZBCP in a vortex decreases with increasing magnetic field, or alternatively, with increasing vortex density \cite{MachidaArX2018}. Therefore, there may be some inherent inter-vortex interactions that can lead to poisoning of MZMs when the vortex pair is brought close enough to fuse. The nature of these interactions can be studied by STM measurement of ZBCPs before and after fusing a vortex pair to gather statistics on the poisoning likelihood. If the fusion process does not lead to high rates of MZM disappearance, then the the qubit readout operation remains viable.

\begin{figure}[h]
   \begin{center}
    \includegraphics[clip=true,width=\columnwidth]{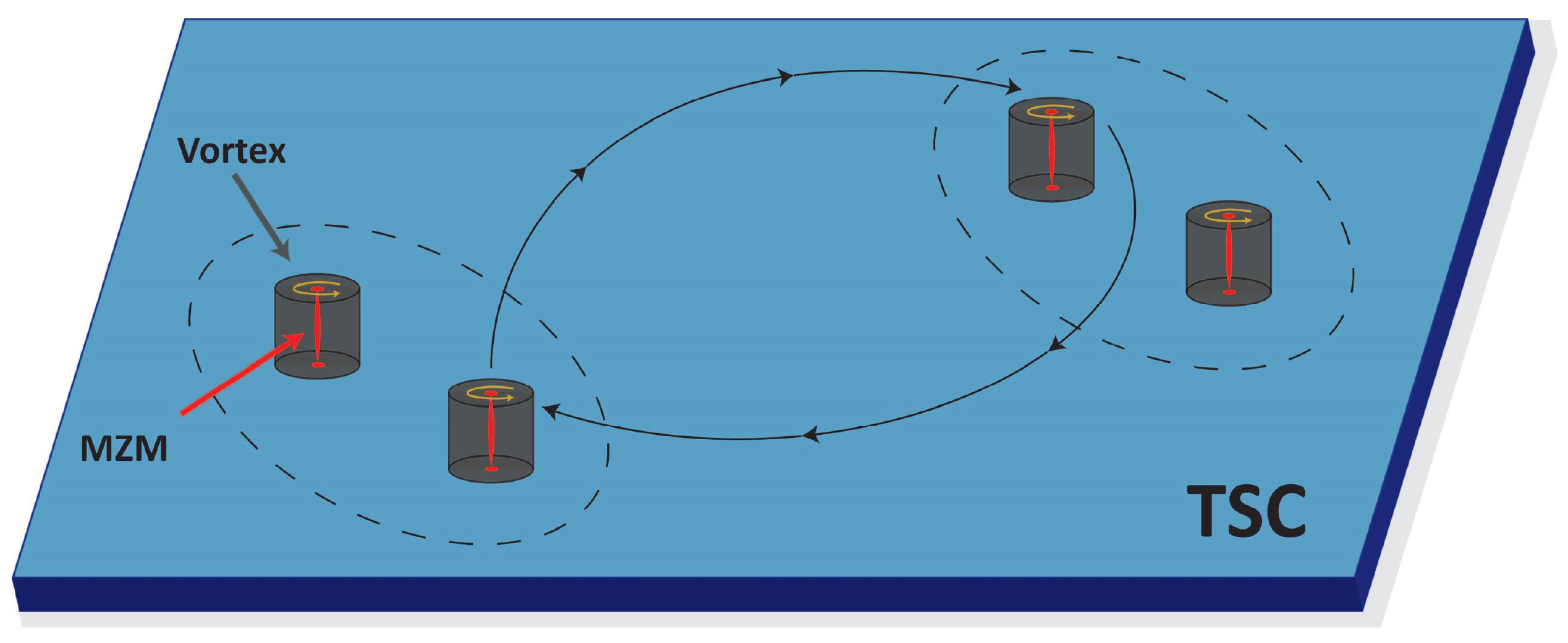}
    \end{center}
    \caption{Diagram of simple qubit braiding algorithm.}
    \label{fig:algorithm}
\end{figure}

With the tools presented above, it is now possible to lay out the groundwork for conducting a qubit operation with two pairs of spatially separated MZM-hosting vortices, schematized in Fig.\ \ref{fig:algorithm}. The states $\ket{0}$ and $\ket{1}$ will be used to identify the eigenvectors of the number operator that defines the canonical fermion formed by the Majorana vortex pair. The state $\ket{1}$ thus corresponds to the presence of a telegraph noise in the force measurement carried out by MFM. The two MZM vortex pairs in the qubit are measured to ensure both begin in the $\ket{0}$ state.  With both qubits in the $\ket{0}$ state, a clockwise exchange is performed between two Majoranas in adjacent qubits as shown in Fig.~\ref{fig:algorithm}. This exchange puts the system in the state $\frac{1}{\sqrt{2}}(\ket{0}_{12}\ket{0}_{34}-\ket{1}_{12}\ket{1}_{34})$. Another clockwise exchange can be performed, which then leaves the system in the state $\ket{1}_{12}\ket{1}_{34}$. Note that at each of the prior steps, measurements of the average occupation of each qubit can be taken to ensure the exchange has preserved the coherence of the two-qubit system. An experimental demonstration that the average occupation numbers are $\frac{1}{2}$ and 1 after the two exchanges, respectively, would prove that the exchange of MZMs within vortex pairs truly constitutes a quantum logic operation.
 \FloatBarrier
\FloatBarrier

In conclusion, we have presented a scheme for the dragging, pinning, and parity readout of a a pair of MZMs in the vortex cores of topological superconductor FeTe$_{1-x}$Se$_x$. We have shown that MFM cantilevers are able to apply the force necessary to fuse two MZMs, and can be sensitive enough to measure the change in supercurrent when the occupation number of the resulting fermionic state is non-zero. Thus, the pathway to an experimental realization of a topological quantum logic operation is plainly laid out.


\begin{thebibliography}{26}%
\makeatletter
\providecommand \@ifxundefined [1]{%
 \@ifx{#1\undefined}
}%
\providecommand \@ifnum [1]{%
 \ifnum #1\expandafter \@firstoftwo
 \else \expandafter \@secondoftwo
 \fi
}%
\providecommand \@ifx [1]{%
 \ifx #1\expandafter \@firstoftwo
 \else \expandafter \@secondoftwo
 \fi
}%
\providecommand \natexlab [1]{#1}%
\providecommand \enquote  [1]{``#1''}%
\providecommand \bibnamefont  [1]{#1}%
\providecommand \bibfnamefont [1]{#1}%
\providecommand \citenamefont [1]{#1}%
\providecommand \href@noop [0]{\@secondoftwo}%
\providecommand \href [0]{\begingroup \@sanitize@url \@href}%
\providecommand \@href[1]{\@@startlink{#1}\@@href}%
\providecommand \@@href[1]{\endgroup#1\@@endlink}%
\providecommand \@sanitize@url [0]{\catcode `\\12\catcode `\$12\catcode
  `\&12\catcode `\#12\catcode `\^12\catcode `\_12\catcode `\%12\relax}%
\providecommand \@@startlink[1]{}%
\providecommand \@@endlink[0]{}%
\providecommand \url  [0]{\begingroup\@sanitize@url \@url }%
\providecommand \@url [1]{\endgroup\@href {#1}{\urlprefix }}%
\providecommand \urlprefix  [0]{URL }%
\providecommand \Eprint [0]{\href }%
\providecommand \doibase [0]{http://dx.doi.org/}%
\providecommand \selectlanguage [0]{\@gobble}%
\providecommand \bibinfo  [0]{\@secondoftwo}%
\providecommand \bibfield  [0]{\@secondoftwo}%
\providecommand \translation [1]{[#1]}%
\providecommand \BibitemOpen [0]{}%
\providecommand \bibitemStop [0]{}%
\providecommand \bibitemNoStop [0]{.\EOS\space}%
\providecommand \EOS [0]{\spacefactor3000\relax}%
\providecommand \BibitemShut  [1]{\csname bibitem#1\endcsname}%
\let\auto@bib@innerbib\@empty
\bibitem [{\citenamefont {Kitaev}(2003)}]{KitaevAnnPhys2003}%
  \BibitemOpen
  \bibfield  {author} {\bibinfo {author} {\bibfnamefont {A.Yu.}\ \bibnamefont
  {Kitaev}},\ }\bibfield  {title} {\enquote {\bibinfo {title} {Fault-tolerant
  quantum computation by anyons},}\ }\href {\doibase
  10.1016/S0003-4916(02)00018-0} {\bibfield  {journal} {\bibinfo  {journal}
  {Annals of Physics}\ }\textbf {\bibinfo {volume} {303}},\ \bibinfo {pages}
  {2--30} (\bibinfo {year} {2003})}\BibitemShut {NoStop}%
\bibitem [{\citenamefont {Albrecht}\ \emph {et~al.}()\citenamefont {Albrecht},
  \citenamefont {Higginbotham}, \citenamefont {Madsen}, \citenamefont
  {Kuemmeth}, \citenamefont {Jespersen}, \citenamefont {Nyg{\aa}rd},
  \citenamefont {Krogstrup},\ and\ \citenamefont {Marcus}}]{AlbrechtNat2016}%
  \BibitemOpen
  \bibfield  {author} {\bibinfo {author} {\bibfnamefont {S~M}\ \bibnamefont
  {Albrecht}}, \bibinfo {author} {\bibfnamefont {A~P}\ \bibnamefont
  {Higginbotham}}, \bibinfo {author} {\bibfnamefont {M}~\bibnamefont {Madsen}},
  \bibinfo {author} {\bibfnamefont {F}~\bibnamefont {Kuemmeth}}, \bibinfo
  {author} {\bibfnamefont {T~S}\ \bibnamefont {Jespersen}}, \bibinfo {author}
  {\bibfnamefont {J}~\bibnamefont {Nyg{\aa}rd}}, \bibinfo {author}
  {\bibfnamefont {P}~\bibnamefont {Krogstrup}}, \ and\ \bibinfo {author}
  {\bibfnamefont {C~M}\ \bibnamefont {Marcus}},\ }\bibfield  {title} {\enquote
  {\bibinfo {title} {Exponential protection of zero modes in {M}ajorana
  islands},}\ }\href {\doibase 10.1038/nature17162} {\bibinfo  {journal}
  {Nature}\ ,\ \bibinfo {pages} {206--209}}\BibitemShut {NoStop}%
\bibitem [{\citenamefont {Suominen}\ \emph {et~al.}(2017)\citenamefont
  {Suominen}, \citenamefont {Kjaergaard}, \citenamefont {Hamilton},
  \citenamefont {Shabani}, \citenamefont {Palmstr{\o}m}, \citenamefont
  {Marcus},\ and\ \citenamefont {Nichele}}]{SuominenPRL2017}%
  \BibitemOpen
\bibfield  {journal} {  }\bibfield  {author} {\bibinfo {author} {\bibfnamefont
  {H.~J.}\ \bibnamefont {Suominen}}, \bibinfo {author} {\bibfnamefont
  {M}~\bibnamefont {Kjaergaard}}, \bibinfo {author} {\bibfnamefont {A.~R.}\
  \bibnamefont {Hamilton}}, \bibinfo {author} {\bibfnamefont {J}~\bibnamefont
  {Shabani}}, \bibinfo {author} {\bibfnamefont {C.~J.}\ \bibnamefont
  {Palmstr{\o}m}}, \bibinfo {author} {\bibfnamefont {C.~M.}\ \bibnamefont
  {Marcus}}, \ and\ \bibinfo {author} {\bibfnamefont {F}~\bibnamefont
  {Nichele}},\ }\bibfield  {title} {\enquote {\bibinfo {title} {Zero-energy
  modes from coalescing {A}ndreev states in a two-dimensional
  semiconductor-superconductor hybrid platform},}\ }\href {\doibase
  10.1103/PhysRevLett.119.176805} {\bibfield  {journal} {\bibinfo  {journal}
  {Physical Review Letters}\ }\textbf {\bibinfo {volume} {119}},\ \bibinfo
  {pages} {176805} (\bibinfo {year} {2017})}\BibitemShut {NoStop}%
\bibitem [{\citenamefont {Zhang}\ \emph
  {et~al.}(2018{\natexlab{a}})\citenamefont {Zhang}, \citenamefont {Liu},
  \citenamefont {Gazibegovic}, \citenamefont {Xu}, \citenamefont {Logan},
  \citenamefont {Wang}, \citenamefont {van Loo}, \citenamefont {Bommer},
  \citenamefont {de~Moor}, \citenamefont {Car}, \citenamefont {{Op het Veld}},
  \citenamefont {van Veldhoven}, \citenamefont {Koelling}, \citenamefont
  {Verheijen}, \citenamefont {Pendharkar}, \citenamefont {Pennachio},
  \citenamefont {Shojaei}, \citenamefont {Lee}, \citenamefont {Palmstr{\o}m},
  \citenamefont {Bakkers}, \citenamefont {Sarma},\ and\ \citenamefont
  {Kouwenhoven}}]{ZhangNat2018}%
  \BibitemOpen
  \bibfield  {author} {\bibinfo {author} {\bibfnamefont {Hao}\ \bibnamefont
  {Zhang}}, \bibinfo {author} {\bibfnamefont {Chun-Xiao}\ \bibnamefont {Liu}},
  \bibinfo {author} {\bibfnamefont {Sasa}\ \bibnamefont {Gazibegovic}},
  \bibinfo {author} {\bibfnamefont {Di}~\bibnamefont {Xu}}, \bibinfo {author}
  {\bibfnamefont {John~A}\ \bibnamefont {Logan}}, \bibinfo {author}
  {\bibfnamefont {Guanzhong}\ \bibnamefont {Wang}}, \bibinfo {author}
  {\bibfnamefont {Nick}\ \bibnamefont {van Loo}}, \bibinfo {author}
  {\bibfnamefont {Jouri D~S}\ \bibnamefont {Bommer}}, \bibinfo {author}
  {\bibfnamefont {Michiel W.~A.}\ \bibnamefont {de~Moor}}, \bibinfo {author}
  {\bibfnamefont {Diana}\ \bibnamefont {Car}}, \bibinfo {author} {\bibfnamefont
  {Roy L.~M.}\ \bibnamefont {{Op het Veld}}}, \bibinfo {author} {\bibfnamefont
  {Petrus~J.}\ \bibnamefont {van Veldhoven}}, \bibinfo {author} {\bibfnamefont
  {Sebastian}\ \bibnamefont {Koelling}}, \bibinfo {author} {\bibfnamefont
  {Marcel~A}\ \bibnamefont {Verheijen}}, \bibinfo {author} {\bibfnamefont
  {Mihir}\ \bibnamefont {Pendharkar}}, \bibinfo {author} {\bibfnamefont
  {Daniel~J}\ \bibnamefont {Pennachio}}, \bibinfo {author} {\bibfnamefont
  {Borzoyeh}\ \bibnamefont {Shojaei}}, \bibinfo {author} {\bibfnamefont
  {Joon~Sue}\ \bibnamefont {Lee}}, \bibinfo {author} {\bibfnamefont {Chris~J}\
  \bibnamefont {Palmstr{\o}m}}, \bibinfo {author} {\bibfnamefont {Erik P A~M}\
  \bibnamefont {Bakkers}}, \bibinfo {author} {\bibfnamefont {S.~Das}\
  \bibnamefont {Sarma}}, \ and\ \bibinfo {author} {\bibfnamefont {Leo~P}\
  \bibnamefont {Kouwenhoven}},\ }\bibfield  {title} {\enquote {\bibinfo {title}
  {Quantized {M}ajorana conductance},}\ }\href {\doibase 10.1038/nature26142}
  {\bibfield  {journal} {\bibinfo  {journal} {Nature}\ }\textbf {\bibinfo
  {volume} {556}},\ \bibinfo {pages} {74--79} (\bibinfo {year}
  {2018}{\natexlab{a}})}\BibitemShut {NoStop}%
\bibitem [{\citenamefont {Liu}\ \emph {et~al.}(2017)\citenamefont {Liu},
  \citenamefont {Sau}, \citenamefont {Stanescu},\ and\ \citenamefont {{Das
  Sarma}}}]{LiuPRB2017}%
  \BibitemOpen
  \bibfield  {author} {\bibinfo {author} {\bibfnamefont {Chun-Xiao}\
  \bibnamefont {Liu}}, \bibinfo {author} {\bibfnamefont {Jay~D}\ \bibnamefont
  {Sau}}, \bibinfo {author} {\bibfnamefont {Tudor~D}\ \bibnamefont {Stanescu}},
  \ and\ \bibinfo {author} {\bibfnamefont {S}~\bibnamefont {{Das Sarma}}},\
  }\bibfield  {title} {\enquote {\bibinfo {title} {{A}ndreev bound states
  versus {M}ajorana bound states in quantum dot-nanowire-superconductor hybrid
  structures: Trivial versus topological zero-bias conductance peaks},}\ }\href
  {\doibase 10.1103/PhysRevB.96.075161} {\bibfield  {journal} {\bibinfo
  {journal} {Physical Review B}\ }\textbf {\bibinfo {volume} {96}},\ \bibinfo
  {pages} {075161} (\bibinfo {year} {2017})}\BibitemShut {NoStop}%
\bibitem [{\citenamefont {Potter}\ and\ \citenamefont
  {Lee}(2011)}]{PotterPRB2011}%
  \BibitemOpen
  \bibfield  {author} {\bibinfo {author} {\bibfnamefont {Andrew~C}\
  \bibnamefont {Potter}}\ and\ \bibinfo {author} {\bibfnamefont {Patrick~A}\
  \bibnamefont {Lee}},\ }\bibfield  {title} {\enquote {\bibinfo {title}
  {Engineering a $p+ip$ superconductor: Comparison of topological insulator and
  {R}ashba spin-orbit-coupled materials},}\ }\href {\doibase
  10.1103/PhysRevB.83.184520} {\bibfield  {journal} {\bibinfo  {journal}
  {Physical Review B}\ }\textbf {\bibinfo {volume} {83}},\ \bibinfo {pages}
  {184520} (\bibinfo {year} {2011})}\BibitemShut {NoStop}%
\bibitem [{\citenamefont {Fu}\ and\ \citenamefont {Kane}(2008)}]{FuPRL2008}%
  \BibitemOpen
  \bibfield  {author} {\bibinfo {author} {\bibfnamefont {Liang}\ \bibnamefont
  {Fu}}\ and\ \bibinfo {author} {\bibfnamefont {C~L}\ \bibnamefont {Kane}},\
  }\bibfield  {title} {\enquote {\bibinfo {title} {{Superconducting Proximity
  Effect and Majorana Fermions at the Surface of a Topological Insulator}},}\
  }\href {\doibase 10.1103/PhysRevLett.100.096407} {\bibfield  {journal}
  {\bibinfo  {journal} {Physical Review Letters}\ }\textbf {\bibinfo {volume}
  {100}},\ \bibinfo {pages} {096407} (\bibinfo {year} {2008})}\BibitemShut
  {NoStop}%
\bibitem [{\citenamefont {Xu}\ \emph {et~al.}(2016)\citenamefont {Xu},
  \citenamefont {Lian}, \citenamefont {Tang}, \citenamefont {Qi},\ and\
  \citenamefont {Zhang}}]{XuPRL2016}%
  \BibitemOpen
  \bibfield  {author} {\bibinfo {author} {\bibfnamefont {Gang}\ \bibnamefont
  {Xu}}, \bibinfo {author} {\bibfnamefont {Biao}\ \bibnamefont {Lian}},
  \bibinfo {author} {\bibfnamefont {Peizhe}\ \bibnamefont {Tang}}, \bibinfo
  {author} {\bibfnamefont {Xiao-Liang}\ \bibnamefont {Qi}}, \ and\ \bibinfo
  {author} {\bibfnamefont {Shou-Cheng}\ \bibnamefont {Zhang}},\ }\bibfield
  {title} {\enquote {\bibinfo {title} {Topological superconductivity on the
  surface of {F}e-based superconductors},}\ }\href {\doibase
  10.1103/PhysRevLett.117.047001} {\bibfield  {journal} {\bibinfo  {journal}
  {Physical Review Letters}\ }\textbf {\bibinfo {volume} {117}},\ \bibinfo
  {pages} {047001} (\bibinfo {year} {2016})}\BibitemShut {NoStop}%
\bibitem [{\citenamefont {Zhang}\ \emph {et~al.}(2019)\citenamefont {Zhang},
  \citenamefont {Wang}, \citenamefont {Wu}, \citenamefont {Yaji}, \citenamefont
  {Ishida}, \citenamefont {Kohama}, \citenamefont {Dai}, \citenamefont {Sun},
  \citenamefont {Bareille}, \citenamefont {Kuroda}, \citenamefont {Kondo},
  \citenamefont {Okazaki}, \citenamefont {Kindo}, \citenamefont {Wang},
  \citenamefont {Jin}, \citenamefont {Hu}, \citenamefont {Thomale},
  \citenamefont {Sumida}, \citenamefont {Wu}, \citenamefont {Miyamoto},
  \citenamefont {Okuda}, \citenamefont {Ding}, \citenamefont {Gu},
  \citenamefont {Tamegai}, \citenamefont {Kawakami}, \citenamefont {Sato},\
  and\ \citenamefont {Shin}}]{ZhangNatPhys2019}%
  \BibitemOpen
  \bibfield  {author} {\bibinfo {author} {\bibfnamefont {Peng}\ \bibnamefont
  {Zhang}}, \bibinfo {author} {\bibfnamefont {Zhijun}\ \bibnamefont {Wang}},
  \bibinfo {author} {\bibfnamefont {Xianxin}\ \bibnamefont {Wu}}, \bibinfo
  {author} {\bibfnamefont {Koichiro}\ \bibnamefont {Yaji}}, \bibinfo {author}
  {\bibfnamefont {Yukiaki}\ \bibnamefont {Ishida}}, \bibinfo {author}
  {\bibfnamefont {Yoshimitsu}\ \bibnamefont {Kohama}}, \bibinfo {author}
  {\bibfnamefont {Guangyang}\ \bibnamefont {Dai}}, \bibinfo {author}
  {\bibfnamefont {Yue}\ \bibnamefont {Sun}}, \bibinfo {author} {\bibfnamefont
  {Cedric}\ \bibnamefont {Bareille}}, \bibinfo {author} {\bibfnamefont {Kenta}\
  \bibnamefont {Kuroda}}, \bibinfo {author} {\bibfnamefont {Takeshi}\
  \bibnamefont {Kondo}}, \bibinfo {author} {\bibfnamefont {Kozo}\ \bibnamefont
  {Okazaki}}, \bibinfo {author} {\bibfnamefont {Koichi}\ \bibnamefont {Kindo}},
  \bibinfo {author} {\bibfnamefont {Xiancheng}\ \bibnamefont {Wang}}, \bibinfo
  {author} {\bibfnamefont {Changqing}\ \bibnamefont {Jin}}, \bibinfo {author}
  {\bibfnamefont {Jiangping}\ \bibnamefont {Hu}}, \bibinfo {author}
  {\bibfnamefont {Ronny}\ \bibnamefont {Thomale}}, \bibinfo {author}
  {\bibfnamefont {Kazuki}\ \bibnamefont {Sumida}}, \bibinfo {author}
  {\bibfnamefont {Shilong}\ \bibnamefont {Wu}}, \bibinfo {author}
  {\bibfnamefont {Koji}\ \bibnamefont {Miyamoto}}, \bibinfo {author}
  {\bibfnamefont {Taichi}\ \bibnamefont {Okuda}}, \bibinfo {author}
  {\bibfnamefont {Hong}\ \bibnamefont {Ding}}, \bibinfo {author} {\bibfnamefont
  {G~D}\ \bibnamefont {Gu}}, \bibinfo {author} {\bibfnamefont {Tsuyoshi}\
  \bibnamefont {Tamegai}}, \bibinfo {author} {\bibfnamefont {Takuto}\
  \bibnamefont {Kawakami}}, \bibinfo {author} {\bibfnamefont {Masatoshi}\
  \bibnamefont {Sato}}, \ and\ \bibinfo {author} {\bibfnamefont {Shik}\
  \bibnamefont {Shin}},\ }\bibfield  {title} {\enquote {\bibinfo {title}
  {Multiple topological states in iron-based superconductors},}\ }\href
  {\doibase 10.1038/s41567-018-0280-z} {\bibfield  {journal} {\bibinfo
  {journal} {Nature Physics}\ }\textbf {\bibinfo {volume} {15}},\ \bibinfo
  {pages} {41--47} (\bibinfo {year} {2019})}\BibitemShut {NoStop}%
\bibitem [{\citenamefont {Zhang}\ \emph
  {et~al.}(2018{\natexlab{b}})\citenamefont {Zhang}, \citenamefont {Yaji},
  \citenamefont {Hashimoto}, \citenamefont {Ota}, \citenamefont {Kondo},
  \citenamefont {Okazaki}, \citenamefont {Wang}, \citenamefont {Wen},
  \citenamefont {Gu}, \citenamefont {Ding},\ and\ \citenamefont
  {Shin}}]{ZhangSci2018}%
  \BibitemOpen
  \bibfield  {author} {\bibinfo {author} {\bibfnamefont {Peng}\ \bibnamefont
  {Zhang}}, \bibinfo {author} {\bibfnamefont {Koichiro}\ \bibnamefont {Yaji}},
  \bibinfo {author} {\bibfnamefont {Takahiro}\ \bibnamefont {Hashimoto}},
  \bibinfo {author} {\bibfnamefont {Yuichi}\ \bibnamefont {Ota}}, \bibinfo
  {author} {\bibfnamefont {Takeshi}\ \bibnamefont {Kondo}}, \bibinfo {author}
  {\bibfnamefont {Kozo}\ \bibnamefont {Okazaki}}, \bibinfo {author}
  {\bibfnamefont {Zhijun}\ \bibnamefont {Wang}}, \bibinfo {author}
  {\bibfnamefont {Jinsheng}\ \bibnamefont {Wen}}, \bibinfo {author}
  {\bibfnamefont {G~D}\ \bibnamefont {Gu}}, \bibinfo {author} {\bibfnamefont
  {Hong}\ \bibnamefont {Ding}}, \ and\ \bibinfo {author} {\bibfnamefont {Shik}\
  \bibnamefont {Shin}},\ }\bibfield  {title} {\enquote {\bibinfo {title}
  {Observation of topological superconductivity on the surface of an iron-based
  superconductor},}\ }\href {\doibase 10.1126/science.aan4596} {\bibfield
  {journal} {\bibinfo  {journal} {Science}\ }\textbf {\bibinfo {volume}
  {360}},\ \bibinfo {pages} {182--186} (\bibinfo {year}
  {2018}{\natexlab{b}})}\BibitemShut {NoStop}%
\bibitem [{\citenamefont {Caroli}\ \emph {et~al.}(1964)\citenamefont {Caroli},
  \citenamefont {{De Gennes}},\ and\ \citenamefont {Matricon}}]{CaroliJPL1964}%
  \BibitemOpen
  \bibfield  {author} {\bibinfo {author} {\bibfnamefont {C}~\bibnamefont
  {Caroli}}, \bibinfo {author} {\bibfnamefont {P~G}\ \bibnamefont {{De
  Gennes}}}, \ and\ \bibinfo {author} {\bibfnamefont {J}~\bibnamefont
  {Matricon}},\ }\href
  {https://pdf.sciencedirectassets.com/272686/1-s2.0-S0031916300X00625/1-s2.0-0031916364903750/main.pdf?x-amz-security-token=AgoJb3JpZ2luX2VjEHMaCXVzLWVhc3QtMSJHMEUCIQCOSxyNDrcAEoFOmZpCUfYzdYVYxcfM{\%}2FMhS9Ejm5xiLIAIgMyHfLd4iGzeW24vDjE2vlt2oxtJ87rn4o97N4FrC9ts}
  {\emph {\bibinfo {title} {Bound Fermion States on a Vortex Line in a Type II
  Superconductor}}},\ \bibinfo {type} {Tech. Rep.}\ (\bibinfo {year}
  {1964})\BibitemShut {NoStop}%
\bibitem [{\citenamefont {Liu}\ \emph {et~al.}(2018)\citenamefont {Liu},
  \citenamefont {Chen}, \citenamefont {Zhang}, \citenamefont {Peng},
  \citenamefont {Yan}, \citenamefont {Wen}, \citenamefont {Lou}, \citenamefont
  {Huang}, \citenamefont {Tian}, \citenamefont {Dong}, \citenamefont {Wang},
  \citenamefont {Bao}, \citenamefont {Wang}, \citenamefont {Yin}, \citenamefont
  {Zhao},\ and\ \citenamefont {Feng}}]{LiuPRX2018}%
  \BibitemOpen
  \bibfield  {author} {\bibinfo {author} {\bibfnamefont {Qin}\ \bibnamefont
  {Liu}}, \bibinfo {author} {\bibfnamefont {Chen}\ \bibnamefont {Chen}},
  \bibinfo {author} {\bibfnamefont {Tong}\ \bibnamefont {Zhang}}, \bibinfo
  {author} {\bibfnamefont {Rui}\ \bibnamefont {Peng}}, \bibinfo {author}
  {\bibfnamefont {Ya-Jun}\ \bibnamefont {Yan}}, \bibinfo {author}
  {\bibfnamefont {Chen-Hao-Ping}\ \bibnamefont {Wen}}, \bibinfo {author}
  {\bibfnamefont {Xia}\ \bibnamefont {Lou}}, \bibinfo {author} {\bibfnamefont
  {Yu-Long}\ \bibnamefont {Huang}}, \bibinfo {author} {\bibfnamefont
  {Jin-Peng}\ \bibnamefont {Tian}}, \bibinfo {author} {\bibfnamefont {Xiao-Li}\
  \bibnamefont {Dong}}, \bibinfo {author} {\bibfnamefont {Guang-Wei}\
  \bibnamefont {Wang}}, \bibinfo {author} {\bibfnamefont {Wei-Cheng}\
  \bibnamefont {Bao}}, \bibinfo {author} {\bibfnamefont {Qiang-Hua}\
  \bibnamefont {Wang}}, \bibinfo {author} {\bibfnamefont {Zhi-Ping}\
  \bibnamefont {Yin}}, \bibinfo {author} {\bibfnamefont {Zhong-Xian}\
  \bibnamefont {Zhao}}, \ and\ \bibinfo {author} {\bibfnamefont {Dong-Lai}\
  \bibnamefont {Feng}},\ }\bibfield  {title} {\enquote {\bibinfo {title}
  {Robust and clean {M}ajorana zero mode in the vortex core of high-temperature
  superconductor {(Li$_{0.84}$Fe$_{0.16}$)OHFeSe}},}\ }\href {\doibase
  10.1103/PhysRevX.8.041056} {\bibfield  {journal} {\bibinfo  {journal}
  {Physical Review X}\ }\textbf {\bibinfo {volume} {8}},\ \bibinfo {pages}
  {041056} (\bibinfo {year} {2018})}\BibitemShut {NoStop}%
\bibitem [{\citenamefont {K{\"{o}}nig}\ and\ \citenamefont
  {Coleman}()}]{KonigArX2019}%
  \BibitemOpen
  \bibfield  {author} {\bibinfo {author} {\bibfnamefont {Elio~J}\ \bibnamefont
  {K{\"{o}}nig}}\ and\ \bibinfo {author} {\bibfnamefont {Piers}\ \bibnamefont
  {Coleman}},\ }\bibfield  {title} {\enquote {\bibinfo {title} {Helical
  {M}ajorana modes in iron based {D}irac superconductors},}\ }\href@noop {} {\
  }\Eprint {http://arxiv.org/abs/1901.03692} {arXiv:1901.03692} \BibitemShut
  {NoStop}%
\bibitem [{\citenamefont {Hao}\ and\ \citenamefont {Hu}(2018)}]{HaoArX2018}%
  \BibitemOpen
  \bibfield  {author} {\bibinfo {author} {\bibfnamefont {Ning}\ \bibnamefont
  {Hao}}\ and\ \bibinfo {author} {\bibfnamefont {Jiangping}\ \bibnamefont
  {Hu}},\ }\bibfield  {title} {\enquote {\bibinfo {title} {Topological quantum
  states of matter in iron-based superconductors: From concepts to material
  realization},}\ }\href@noop {} {\  (\bibinfo {year} {2018})},\ \Eprint
  {http://arxiv.org/abs/1811.03802} {arXiv:1811.03802} \BibitemShut {NoStop}%
\bibitem [{\citenamefont {Wang}\ \emph {et~al.}(2018)\citenamefont {Wang},
  \citenamefont {Kong}, \citenamefont {Fan}, \citenamefont {Chen},
  \citenamefont {Zhu}, \citenamefont {Liu}, \citenamefont {Cao}, \citenamefont
  {Sun}, \citenamefont {Du}, \citenamefont {Schneeloch}, \citenamefont {Zhong},
  \citenamefont {Gu}, \citenamefont {Fu}, \citenamefont {Ding},\ and\
  \citenamefont {Gao}}]{WangSci2018}%
  \BibitemOpen
  \bibfield  {author} {\bibinfo {author} {\bibfnamefont {Dongfei}\ \bibnamefont
  {Wang}}, \bibinfo {author} {\bibfnamefont {Lingyuan}\ \bibnamefont {Kong}},
  \bibinfo {author} {\bibfnamefont {Peng}\ \bibnamefont {Fan}}, \bibinfo
  {author} {\bibfnamefont {Hui}\ \bibnamefont {Chen}}, \bibinfo {author}
  {\bibfnamefont {Shiyu}\ \bibnamefont {Zhu}}, \bibinfo {author} {\bibfnamefont
  {Wenyao}\ \bibnamefont {Liu}}, \bibinfo {author} {\bibfnamefont
  {Lu}~\bibnamefont {Cao}}, \bibinfo {author} {\bibfnamefont {Yujie}\
  \bibnamefont {Sun}}, \bibinfo {author} {\bibfnamefont {Shixuan}\ \bibnamefont
  {Du}}, \bibinfo {author} {\bibfnamefont {John}\ \bibnamefont {Schneeloch}},
  \bibinfo {author} {\bibfnamefont {Ruidan}\ \bibnamefont {Zhong}}, \bibinfo
  {author} {\bibfnamefont {Genda}\ \bibnamefont {Gu}}, \bibinfo {author}
  {\bibfnamefont {Liang}\ \bibnamefont {Fu}}, \bibinfo {author} {\bibfnamefont
  {Hong}\ \bibnamefont {Ding}}, \ and\ \bibinfo {author} {\bibfnamefont
  {Hong~Jun}\ \bibnamefont {Gao}},\ }\bibfield  {title} {\enquote {\bibinfo
  {title} {Evidence for {M}ajorana bound states in an iron-based
  superconductor},}\ }\href {\doibase 10.1126/science.aao1797} {\bibfield
  {journal} {\bibinfo  {journal} {Science}\ }\textbf {\bibinfo {volume}
  {362}},\ \bibinfo {pages} {333--335} (\bibinfo {year} {2018})}\BibitemShut
  {NoStop}%
\bibitem [{\citenamefont {Machida}\ \emph {et~al.}(2018)\citenamefont
  {Machida}, \citenamefont {Sun}, \citenamefont {Pyon}, \citenamefont {Takeda},
  \citenamefont {Kohsaka}, \citenamefont {Hanaguri}, \citenamefont {Sasagawa},\
  and\ \citenamefont {Tamegai}}]{MachidaArX2018}%
  \BibitemOpen
  \bibfield  {author} {\bibinfo {author} {\bibfnamefont {T}~\bibnamefont
  {Machida}}, \bibinfo {author} {\bibfnamefont {Y}~\bibnamefont {Sun}},
  \bibinfo {author} {\bibfnamefont {S}~\bibnamefont {Pyon}}, \bibinfo {author}
  {\bibfnamefont {S}~\bibnamefont {Takeda}}, \bibinfo {author} {\bibfnamefont
  {Y}~\bibnamefont {Kohsaka}}, \bibinfo {author} {\bibfnamefont
  {T}~\bibnamefont {Hanaguri}}, \bibinfo {author} {\bibfnamefont
  {T}~\bibnamefont {Sasagawa}}, \ and\ \bibinfo {author} {\bibfnamefont
  {T}~\bibnamefont {Tamegai}},\ }\bibfield  {title} {\enquote {\bibinfo {title}
  {Zero-energy vortex bound state in the superconducting topological surface
  state of {Fe(Se,Te)}},}\ }\href@noop {} {\  (\bibinfo {year} {2018})},\
  \Eprint {http://arxiv.org/abs/1812.08995} {arXiv:1812.08995} \BibitemShut
  {NoStop}%
\bibitem [{\citenamefont {Jackiw}\ and\ \citenamefont
  {Rossi}(1981)}]{JackiwNucPhysB1981}%
  \BibitemOpen
  \bibfield  {author} {\bibinfo {author} {\bibfnamefont {R.}~\bibnamefont
  {Jackiw}}\ and\ \bibinfo {author} {\bibfnamefont {P.}~\bibnamefont {Rossi}},\
  }\bibfield  {title} {\enquote {\bibinfo {title} {Zero modes of the
  vortex-fermion system},}\ }\href {\doibase 10.1016/0550-3213(81)90044-4}
  {\bibfield  {journal} {\bibinfo  {journal} {Nuclear Physics B}\ }\textbf
  {\bibinfo {volume} {190}},\ \bibinfo {pages} {681--691} (\bibinfo {year}
  {1981})}\BibitemShut {NoStop}%
\bibitem [{\citenamefont {Straver}\ \emph {et~al.}(2008)\citenamefont
  {Straver}, \citenamefont {Hoffman}, \citenamefont {Auslaender}, \citenamefont
  {Rugar},\ and\ \citenamefont {Moler}}]{StraverAPL2008}%
  \BibitemOpen
  \bibfield  {author} {\bibinfo {author} {\bibfnamefont {E.~W.~J.}\
  \bibnamefont {Straver}}, \bibinfo {author} {\bibfnamefont {Jennifer~E.}\
  \bibnamefont {Hoffman}}, \bibinfo {author} {\bibfnamefont {Ophir~M.}\
  \bibnamefont {Auslaender}}, \bibinfo {author} {\bibfnamefont
  {D.}~\bibnamefont {Rugar}}, \ and\ \bibinfo {author} {\bibfnamefont
  {Kathryn~A.}\ \bibnamefont {Moler}},\ }\bibfield  {title} {\enquote {\bibinfo
  {title} {Controlled manipulation of individual vortices in a
  superconductor},}\ }\href {\doibase 10.1063/1.3000963} {\bibfield  {journal}
  {\bibinfo  {journal} {Applied Physics Letters}\ }\textbf {\bibinfo {volume}
  {93}},\ \bibinfo {pages} {172514} (\bibinfo {year} {2008})}\BibitemShut
  {NoStop}%
\bibitem [{\citenamefont {Catelani}\ \emph {et~al.}(2011)\citenamefont
  {Catelani}, \citenamefont {Koch}, \citenamefont {Frunzio}, \citenamefont
  {Schoelkopf}, \citenamefont {Devoret},\ and\ \citenamefont
  {Glazman}}]{CatelaniPRL2011}%
  \BibitemOpen
  \bibfield  {author} {\bibinfo {author} {\bibfnamefont {G}~\bibnamefont
  {Catelani}}, \bibinfo {author} {\bibfnamefont {J}~\bibnamefont {Koch}},
  \bibinfo {author} {\bibfnamefont {L}~\bibnamefont {Frunzio}}, \bibinfo
  {author} {\bibfnamefont {R~J}\ \bibnamefont {Schoelkopf}}, \bibinfo {author}
  {\bibfnamefont {M~H}\ \bibnamefont {Devoret}}, \ and\ \bibinfo {author}
  {\bibfnamefont {L~I}\ \bibnamefont {Glazman}},\ }\bibfield  {title} {\enquote
  {\bibinfo {title} {Quasiparticle relaxation of superconducting qubits in the
  presence of flux},}\ }\href {\doibase 10.1103/PhysRevLett.106.077002}
  {\bibfield  {journal} {\bibinfo  {journal} {Physical Review Letters}\
  }\textbf {\bibinfo {volume} {106}},\ \bibinfo {pages} {077002} (\bibinfo
  {year} {2011})}\BibitemShut {NoStop}%
\bibitem [{\citenamefont {Albrecht}\ \emph {et~al.}(2017)\citenamefont
  {Albrecht}, \citenamefont {Hansen}, \citenamefont {Higginbotham},
  \citenamefont {Kuemmeth}, \citenamefont {Jespersen}, \citenamefont
  {Nyg{\aa}rd}, \citenamefont {Krogstrup}, \citenamefont {Danon}, \citenamefont
  {Flensberg},\ and\ \citenamefont {Marcus}}]{AlbrechtPRL2017}%
  \BibitemOpen
  \bibfield  {author} {\bibinfo {author} {\bibfnamefont {S.~M.}\ \bibnamefont
  {Albrecht}}, \bibinfo {author} {\bibfnamefont {E.~B.}\ \bibnamefont
  {Hansen}}, \bibinfo {author} {\bibfnamefont {A.~P.}\ \bibnamefont
  {Higginbotham}}, \bibinfo {author} {\bibfnamefont {F}~\bibnamefont
  {Kuemmeth}}, \bibinfo {author} {\bibfnamefont {T.~S.}\ \bibnamefont
  {Jespersen}}, \bibinfo {author} {\bibfnamefont {J}~\bibnamefont
  {Nyg{\aa}rd}}, \bibinfo {author} {\bibfnamefont {P}~\bibnamefont
  {Krogstrup}}, \bibinfo {author} {\bibfnamefont {J}~\bibnamefont {Danon}},
  \bibinfo {author} {\bibfnamefont {K}~\bibnamefont {Flensberg}}, \ and\
  \bibinfo {author} {\bibfnamefont {C.~M.}\ \bibnamefont {Marcus}},\ }\bibfield
   {title} {\enquote {\bibinfo {title} {Transport signatures of quasiparticle
  poisoning in a {M}ajorana island},}\ }\href {\doibase
  10.1103/PhysRevLett.118.137701} {\bibfield  {journal} {\bibinfo  {journal}
  {Physical Review Letters}\ }\textbf {\bibinfo {volume} {118}},\ \bibinfo
  {pages} {137701} (\bibinfo {year} {2017})}\BibitemShut {NoStop}%
\bibitem [{\citenamefont {Chaves}\ \emph {et~al.}(2011)\citenamefont {Chaves},
  \citenamefont {Peeters}, \citenamefont {Farias},\ and\ \citenamefont
  {Milo{\v{s}}evi{\'{c}}}}]{ChavesPRB2011}%
  \BibitemOpen
  \bibfield  {author} {\bibinfo {author} {\bibfnamefont {Andrey}\ \bibnamefont
  {Chaves}}, \bibinfo {author} {\bibfnamefont {F.~M.}\ \bibnamefont {Peeters}},
  \bibinfo {author} {\bibfnamefont {G.~A.}\ \bibnamefont {Farias}}, \ and\
  \bibinfo {author} {\bibfnamefont {M.~V.}\ \bibnamefont
  {Milo{\v{s}}evi{\'{c}}}},\ }\bibfield  {title} {\enquote {\bibinfo {title}
  {Vortex-vortex interaction in bulk superconductors: {G}inzburg-{L}andau
  theory},}\ }\href {\doibase 10.1103/PhysRevB.83.054516} {\bibfield  {journal}
  {\bibinfo  {journal} {Physical Review B}\ }\textbf {\bibinfo {volume} {83}},\
  \bibinfo {pages} {054516} (\bibinfo {year} {2011})}\BibitemShut {NoStop}%
\bibitem [{\citenamefont {Auslaender}\ \emph {et~al.}(2008)\citenamefont
  {Auslaender}, \citenamefont {Luan}, \citenamefont {Straver}, \citenamefont
  {Hoffman}, \citenamefont {Koshnick}, \citenamefont {Zeldov}, \citenamefont
  {Bonn}, \citenamefont {Liang}, \citenamefont {Hardy},\ and\ \citenamefont
  {Moler}}]{AuslaenderNatPhys2008}%
  \BibitemOpen
  \bibfield  {author} {\bibinfo {author} {\bibfnamefont {Ophir~M.}\
  \bibnamefont {Auslaender}}, \bibinfo {author} {\bibfnamefont {Lan}\
  \bibnamefont {Luan}}, \bibinfo {author} {\bibfnamefont {Eric W.~J.}\
  \bibnamefont {Straver}}, \bibinfo {author} {\bibfnamefont {Jennifer~E.}\
  \bibnamefont {Hoffman}}, \bibinfo {author} {\bibfnamefont {Nicholas~C.}\
  \bibnamefont {Koshnick}}, \bibinfo {author} {\bibfnamefont {Eli}\
  \bibnamefont {Zeldov}}, \bibinfo {author} {\bibfnamefont {Douglas~A.}\
  \bibnamefont {Bonn}}, \bibinfo {author} {\bibfnamefont {Ruixing}\
  \bibnamefont {Liang}}, \bibinfo {author} {\bibfnamefont {Walter~N.}\
  \bibnamefont {Hardy}}, \ and\ \bibinfo {author} {\bibfnamefont {Kathryn~A.}\
  \bibnamefont {Moler}},\ }\bibfield  {title} {\enquote {\bibinfo {title}
  {Mechanics of individual isolated vortices in a cuprate superconductor},}\
  }\href {\doibase 10.1038/nphys1127} {\bibfield  {journal} {\bibinfo
  {journal} {Nature Physics}\ }\textbf {\bibinfo {volume} {5}},\ \bibinfo
  {pages} {35--39} (\bibinfo {year} {2008})}\BibitemShut {NoStop}%
\bibitem [{\citenamefont {Zhang}\ \emph {et~al.}(2015)\citenamefont {Zhang},
  \citenamefont {Kim}, \citenamefont {Huefner}, \citenamefont {Ye},
  \citenamefont {Kim}, \citenamefont {Canfield}, \citenamefont {Prozorov},
  \citenamefont {Auslaender},\ and\ \citenamefont {Hoffman}}]{ZhangPRB2015}%
  \BibitemOpen
  \bibfield  {author} {\bibinfo {author} {\bibfnamefont {Jessie~T}\
  \bibnamefont {Zhang}}, \bibinfo {author} {\bibfnamefont {Jeehoon}\
  \bibnamefont {Kim}}, \bibinfo {author} {\bibfnamefont {Magdalena}\
  \bibnamefont {Huefner}}, \bibinfo {author} {\bibfnamefont {Cun}\ \bibnamefont
  {Ye}}, \bibinfo {author} {\bibfnamefont {Stella}\ \bibnamefont {Kim}},
  \bibinfo {author} {\bibfnamefont {Paul~C.}\ \bibnamefont {Canfield}},
  \bibinfo {author} {\bibfnamefont {Ruslan}\ \bibnamefont {Prozorov}}, \bibinfo
  {author} {\bibfnamefont {Ophir~M}\ \bibnamefont {Auslaender}}, \ and\
  \bibinfo {author} {\bibfnamefont {Jennifer~E.}\ \bibnamefont {Hoffman}},\
  }\bibfield  {title} {\enquote {\bibinfo {title} {Single-vortex pinning and
  penetration depth in superconducting {NdFeAsO$_{1-x}$F$_x$}},}\ }\href
  {\doibase 10.1103/PhysRevB.92.134509} {\bibfield  {journal} {\bibinfo
  {journal} {Physical Review B}\ }\textbf {\bibinfo {volume} {92}},\ \bibinfo
  {pages} {134509} (\bibinfo {year} {2015})}\BibitemShut {NoStop}%
\bibitem [{\citenamefont {Huefner}\ \emph {et~al.}(2012)\citenamefont
  {Huefner}, \citenamefont {Pivonka}, \citenamefont {Kim}, \citenamefont {Ye},
  \citenamefont {Blood-Forsythe}, \citenamefont {Zech},\ and\ \citenamefont
  {Hoffman}}]{HuefnerAPL2012}%
  \BibitemOpen
  \bibfield  {author} {\bibinfo {author} {\bibfnamefont {Magdalena}\
  \bibnamefont {Huefner}}, \bibinfo {author} {\bibfnamefont {Adam}\
  \bibnamefont {Pivonka}}, \bibinfo {author} {\bibfnamefont {Jeehoon}\
  \bibnamefont {Kim}}, \bibinfo {author} {\bibfnamefont {Cun}\ \bibnamefont
  {Ye}}, \bibinfo {author} {\bibfnamefont {Martin~A.}\ \bibnamefont
  {Blood-Forsythe}}, \bibinfo {author} {\bibfnamefont {Martin}\ \bibnamefont
  {Zech}}, \ and\ \bibinfo {author} {\bibfnamefont {Jennifer~E.}\ \bibnamefont
  {Hoffman}},\ }\bibfield  {title} {\enquote {\bibinfo {title} {Microcantilever
  {Q} control via capacitive coupling},}\ }\href {\doibase 10.1063/1.4764025}
  {\bibfield  {journal} {\bibinfo  {journal} {Applied Physics Letters}\
  }\textbf {\bibinfo {volume} {101}},\ \bibinfo {pages} {173110} (\bibinfo
  {year} {2012})}\BibitemShut {NoStop}%
\bibitem [{\citenamefont {Seo}\ \emph {et~al.}(2005)\citenamefont {Seo},
  \citenamefont {Cadden-Zimansky},\ and\ \citenamefont
  {Chandrasekhar}}]{SeoAPL2005}%
  \BibitemOpen
  \bibfield  {author} {\bibinfo {author} {\bibfnamefont {Yongho}\ \bibnamefont
  {Seo}}, \bibinfo {author} {\bibfnamefont {Paul}\ \bibnamefont
  {Cadden-Zimansky}}, \ and\ \bibinfo {author} {\bibfnamefont {Venkat}\
  \bibnamefont {Chandrasekhar}},\ }\bibfield  {title} {\enquote {\bibinfo
  {title} {Low-temperature high-resolution magnetic force microscopy using a
  quartz tuning fork},}\ }\href {\doibase 10.1063/1.2037852} {\bibfield
  {journal} {\bibinfo  {journal} {Applied Physics Letters}\ }\textbf {\bibinfo
  {volume} {87}},\ \bibinfo {pages} {103103} (\bibinfo {year}
  {2005})}\BibitemShut {NoStop}%
\bibitem [{\citenamefont {Massee}\ \emph {et~al.}(2015)\citenamefont {Massee},
  \citenamefont {Sprau}, \citenamefont {Wang}, \citenamefont {Davis},
  \citenamefont {Ghigo}, \citenamefont {Gu},\ and\ \citenamefont
  {Kwok}}]{MasseeSciAdv2015}%
  \BibitemOpen
  \bibfield  {author} {\bibinfo {author} {\bibfnamefont {F.}~\bibnamefont
  {Massee}}, \bibinfo {author} {\bibfnamefont {P.~O.}\ \bibnamefont {Sprau}},
  \bibinfo {author} {\bibfnamefont {Y.-L.}\ \bibnamefont {Wang}}, \bibinfo
  {author} {\bibfnamefont {J.~C.~S.}\ \bibnamefont {Davis}}, \bibinfo {author}
  {\bibfnamefont {G.}~\bibnamefont {Ghigo}}, \bibinfo {author} {\bibfnamefont
  {G.~D.}\ \bibnamefont {Gu}}, \ and\ \bibinfo {author} {\bibfnamefont {W.-K.}\
  \bibnamefont {Kwok}},\ }\bibfield  {title} {\enquote {\bibinfo {title}
  {Imaging atomic-scale effects of high-energy ion irradiation on
  superconductivity and vortex pinning in {Fe(Se,Te)}},}\ }\href {\doibase
  10.1126/sciadv.1500033} {\bibfield  {journal} {\bibinfo  {journal} {Science
  Advances}\ }\textbf {\bibinfo {volume} {1}},\ \bibinfo {pages} {e1500033}
  (\bibinfo {year} {2015})}\BibitemShut {NoStop}%
\end{thebibliography}

\section{\label{Appendix}Appendix}
\appendix

\section{Derivation and Solution of Radial BdG Equation}
We begin by recalling Eq.~\ref{polBdG}, the BdG Hamiltonian in polar coordinates, and the fact that it commutes with the total angular momentum operator 
\begin{align}
    J=L+\frac{1}{2}(\sigma_z+n\tau_z), 
\end{align}
where $L=-i\partial_{\theta}$ is the orbital angular momentum operator. This allows us to focus on solutions with eigenvalues $J=m$. The spectrum of excitations of this system is found by solving the eigenvalue problem
\begin{align}
H\Psi_m(r,\theta)=E\Psi_m(r,\theta).
\end{align}
The angular dependence for such states is written as 
\begin{align}
\Psi_m(r,\theta')&=e^{i \theta' L }\Psi_m(r,\theta)|_{\theta=0}\nonumber\\
&=e^{-i \theta' (\sigma_z+n\tau_z-2m)/2 }\Psi_m(r,0).
\end{align}
Applying this transformation, we can isolate the theta dependence of the BdG equation as 
\begin{align}
e^{\frac{i \theta}{2} (\sigma_z+n\tau_z-2 m) }He^{-\frac{i \theta}{2} (\sigma_z+n\tau_z-2 m) }\Psi_m(r,0)=E\Psi_m(r,0),
\end{align}
where 
\begin{align}
H_m&=e^{\frac{i \theta}{2} (\sigma_z+n\tau_z-2m) }He^{-\frac{i \theta}{2} (\sigma_z+n\tau_z-2m) }\nonumber\\
&=-i v\sigma_x\tau_z\partial_r-\tau_z[\mu+i (\sigma_y)r^{-1}-\frac{i}{2}(\sigma_z+n\tau_z-2m)]\nonumber\\&+\Delta(r)\tau_x.
\end{align}
The 1-D radial BdG equation then takes the form
\begin{align}
&[-i v\sigma_x\tau_z\partial_r+\tau_z\{-\mu-i (\sigma_y) r^{-1}(-i/2)(\sigma_z+n\tau_z-2m)\}\nonumber\\
&+\Delta(r)\tau_x-E]\Psi(r)=0,
\end{align}
as before noted in Eq.~\ref{radBdG}. The extension to Eq.~\ref{Psi} follows from setting $v=\Delta_0=1$ and making the rotation $\sigma_x\rightarrow\sigma_y$. Eq.~\ref{Psi} can in principle be solved as an initial value problem from $r=0$ to $r=\infty$. In the limit $r\rightarrow 0$, the $\frac{1}{r}$ term dominates and thus the initial value $\Psi(r\rightarrow 0)$ must satisfy the 
constraint
\begin{align}
&\{1+\sigma_z(n\tau_z-2m)\}\Psi(r\rightarrow 0) = \alpha\Psi(r\rightarrow 0),
\end{align}
where $\alpha\leq 0$.
Let us now focus on the $m=1/2$ angular momentum channel of a double vortex ($n=2$). The initial condition then satisfies 
\begin{align}
&(1-\sigma_z+2\sigma_z\tau_z)\Psi(r\rightarrow 0) = \alpha\Psi(r\rightarrow 0).
\end{align}
This allows  initial conditions  $\Psi_j(r\rightarrow 0)=\Psi_j$ where $\Psi_{j=1,2}$ are the two orthonormal vectors with  $\sigma_z\tau_z=-1$ .
Let us represent  the solutions to the initial value problem defined by Eq.~\ref{Psi} as $\Psi_j(r)$. A general solution matching the boundary conditions as $r\rightarrow 0$ is given by 
\begin{align}
&\Psi(r)=\sum_j c_j \Psi_j(r).\label{smallr}
\end{align} 

Next we consider the boundary conditions at $r\rightarrow\infty$, where Eq.~\ref{Psi} takes the form
\begin{align}
&[\partial_r-i\mu\sigma_y-\tau_y\sigma_y-i E\sigma_y\tau_z]\Psi=0.
\end{align}
In this limit $\Psi(r)=\Psi e^{-z r}$ where convergent solutions require $Re[z]>0$.
Substituting we get 
\begin{align}
&[-i\mu\sigma_y-\tau_y\sigma_y-i E\sigma_y\tau_z]\Psi=z\Psi.
\end{align}
Let us denote the two eigenvectors with the convergent eigenvalues as $\tilde{\Psi}_{j=1,2}$.
We can then define $\tilde{\Psi}_j(r)$ to be the solutions of Eq.~\ref{Psi} with boundary conditions $\tilde{\Psi}_j(r\rightarrow\infty)=\tilde{\Psi}_j e^{-z_j r}$.
A general solution matching the boundary conditions as $r\rightarrow \infty$ is given by 

\begin{align}
&\Psi(r)=\sum_j \tilde{c}_j \tilde{\Psi}_j(r).\label{larger}
\end{align}

For a complete solution the two solutions at small and large $r$, given by Eq.~\ref{smallr} and Eq.~\ref{larger}, respectively, must match at an 
intermediate $r=R$. Since this is a linear condition for 4 component wave-functions with 4 coefficients, such a matching is possible only if the energy $E$ satisfies the condition 
\begin{align}
&M(E)=\begin{array}{cccc}\Big [ \Psi_1(R) & \Psi_2(R) & \tilde{\Psi}_1(R)&\tilde{\Psi}_2(R) \Big] \end{array}=0.
\end{align}

Since Eq.~\ref{Psi} is real, $M(E)$ is real and the solutions can be determined by simple root finding.
The null vector of the matrix provides us with the coefficients of $c_j$, which then allows us to construct $\Psi(r)$ using Eq.~\ref{smallr}.
In practice we can compute $\tilde{\Psi}_j(R)$ by solving Eq.~\ref{Psi} in reverse from some large $R'\gg R$ with an arbitrary initial condition. 
While the resulting solution contains the states with $Re[z]<0$, the amplitudes of such states are exponentially small. It can be shown that this is 
equivalent to considering $\Psi_j(R')$ and setting the projector to zero \textbf{add citation}. Therefore, for practical purposes we solve the 
simpler problem
\begin{align}
&M_1(E)=\Big[(1+\sigma_z\tau_z)\begin{array}{cc}\Psi_1(R') & \Psi_2(R') \Big]\end{array}=0.
\end{align}
As before, the bound state energy is the root in $E$ for which the above matrix has a zero eigenvalue.
The null vector of the matrix provides us with the coefficients of $c_j$, which then allows us to construct $\Psi(r)$ using Eq.~\ref{smallr}.

\section{The Chiral Limit}
The radial equation for the vortex (Eq.~\ref{Psi}) is analytically solvable in the chiral limit i.e. $\mu=0$ and $m=1/2$, in which 
case there is a solution for $E=0$. In this limit Eq.~\ref{Psi} becomes  
\begin{align}
&[-\partial_r- r^{-1}(n\sigma_z \tau_z)/2+\tanh(r)\tau_y\sigma_x]\Psi(r)=0.\label{Psic}
\end{align}
This equation is easy to solve since the comprising matrices ($1,\sigma_z\tau_z$ and $\tau_y\sigma_x$) commute. The solution 
to these equations are then formally written as 
\begin{align}
&\Psi(r)=r^{-(n\sigma_z \tau_z)/2}\cosh(r)^{\tau_y\sigma_x}\Psi_0.
\end{align}
For a double vortex this leads to a unique solution of the form 
\begin{align}
&\Psi(r)=N r \sech(r)\Psi_0,
\end{align}
where $\Psi_0$ is defined by the equation $\sigma_z\tau_z\Psi_0=\tau_y\sigma_x\Psi_0=-\Psi_0$ and $N$ is a normalization constant.
Adding a finite chemical potential $\mu$ doesn't affect the wave-function much at lowest order in perturbation theory, but shifts 
the energy to 
\begin{align}
&E\simeq -\mu\Psi_0^\dagger \tau_z\Psi_0=0.
\end{align}
The current density would be given by 
\begin{align}
&j(r)=v N^2 r^2 \sech^2(r)\Psi_0^\dagger \sigma_y\Psi_0=0.
\end{align}
This means one needs to consider the wave-function contribution of order $\mu$, which is out of the scope of our arguments.

\end{document}